\documentclass[preprint,a4paper,aps,prb,showkeys]{revtex4}
\usepackage{epsfig}
\begin{document}

\title{Structural and dynamical characteristics of \\ mesoscopic  H$^+$[H$_2$O]$_n$ clusters}

\author{Mariano Galvagno,$^1$ Daniel Laria,$^{1,2}$ and 
Javier Rodriguez.$^{1,2,}$\footnote{Corresponding author. E-mail: jarodrig@cnea.gov.ar} }

\vspace{1.5truecm}

\affiliation{
      $^1$Unidad Actividad F\' \i sica. Comisi\'on Nacional de Energ\' \i a At\'omica,  \\
      Avenida Libertador 8250, 1429, Buenos Aires, Argentina. \\
      $^2$Departamento de Qu\' \i mica Inorg\'anica,
      Anal\' \i tica y  Qu\'\i mica-F\' \i sica e INQUIMAE.
      Facultad de Ciencias Exactas y Naturales. \\
      Universidad de Buenos Aires, Ciudad Universitaria, Pabell\'on II, 1428, Buenos Aires,
      Argentina. \\
}

\begin{abstract}
Structural and dynamical characteristics pertaining to the solvation of 
an excess proton in liquid-like nanoclusters of the type [H$_2$O]$_n$ are 
investigated using Molecular Dynamics experiments. Three different aggregate 
sizes were analyzed: $n=10,$ 21 and 125. The simulation experiments were 
performed using a multistate empirical valence bond Hamiltonian model. 
While in the smallest aggregates the proton occupies a central position, 
the stable solvation environments for $n=21$ and 125 are located at the cluster 
boundaries. In all cases, the structure of the closest solvation shell of the excess 
charge remains practically unchanged and coincides with that observed in bulk water. 
Compared to results obtained in bulk, the computed rates for proton transfer in
clusters are between one and two orders of magnitude slower, and tend to increase 
for larger cluster sizes.
\end{abstract}

\maketitle

\section{Introduction}
The study of the solvation of ionic species in aqueous clusters is of 
great importance in many areas of environmental and atmospheric 
chemistry.\cite{envi1,envi2,envi3,envi4} Many chemical reactions occurring in the earth 
atmosphere involve ions embedded at the surface of icicles. Protonated 
water clusters are perhaps one of the most abundant of these species
in the stratosphere and can host a large variety of chemical processes;
most notably are those controlling atmospheric nucleation.  For example, 
Yang et al.\cite{yang} have analyzed the importance of these aggregates 
in the earliest stages of the formation of noctilucent clouds at high altitudes, 
at temperatures as low as 150 K. From a more fundamental perspective, the analysis of 
the equilibrium and dynamical characteristics of chemical reactivity in 
these clusters continues to draw considerable attention from experimental 
and theoretical perspectives.\cite{rev1}  One of the reasons for 
this interest is the fact that, by a careful control of the cluster sizes, 
one can bridge the gap between analyses of reactivity in two important 
limiting behaviors, $i.e.$ gaseous and condensed phases.

In the present paper we examine the solvation of an excess proton in mesoscopic 
clusters of the type H$^+$[H$_2$O]$_n$, with sizes intermediate between $n=10$ 
and $n=125$.  This size range covers aggregates with incipient three dimensional 
structures up to nanoclusters where it is possible to establish a clear distinction 
between  inner, ``bulk-like", domains and surface states. Our analysis will 
be based on results obtained from Molecular Dynamics simulation experiments 
and includes the study of the equilibrium solvation structures and dynamical aspects 
related to  proton transfer processes as well.  To tackle the problem we 
resorted on an multistate empirical valence bond (MS-EVB) model 
Hamiltonian.\cite{warshel,accounts,borgis1,borgis2,tuck9} 
This approach has been repeatedly used in the past and has proved to 
be sufficiently versatile so as to be successfully applied for the 
examination of aqueous protons in a wide variety 
of environments: bulk,\cite{borgis1,borgis2,tuck9,voth1,voth2,voth3,voth4,voth5,bulk_k,bulk_k_1}
clusters,\cite{cl,cl1} confined 
environments,\cite{confined1,nafion,voth_cc} air-water interfaces,\cite{interfaces}
and water at extreme conditions of temperature and pressure,\cite{scw}
to cite a just a few important examples.
The organization of the paper is as follows: in Section II we present a brief
overview of the model and the simulation procedure. The main results are 
presented in Section III. The last section contains the concluding remarks.

\section{Model and Simulation Procedure}

The systems under investigation consisted of protonated clusters of the type 
H$^+$[H$_2$O]$_n$, with $n=10$, 21 and 125. The Molecular Dynamics experiments 
were performed using an MS-EVB scheme. This model has been thoroughly described 
in previous studies, so we will present here a brief overview of its main features 
and refer the interested reader to Refs. \cite {voth1,voth2,voth3,voth4,voth5}, 
for more complete presentations.

The cornerstone of the  model is the following MS-EVB Hamiltonian:
\begin{equation}
\label{eq1}
\hat{\rm H}_{{\rm EVB}}(\{{\bf R}\})= \sum_{ij} |\phi_i\rangle h_{ij}(\{{\bf 
R}\})
\langle \phi_j|  \ \ \ ;
\end{equation}
where $\{|\phi_i\rangle\}$ represents a basis set of diabatic states. Each of
these states denotes the spatial localization of the excess proton in a different
water molecule. The matrix elements of the MS-EVB Hamiltonian,
$h_{ij}(\{{\bf R}\})$, are parametrized in terms of the nuclear coordinates 
and are adjusted so that EVB predictions for the energetics and geometrical 
details of small protonated water clusters agree with those obtained from 
highly sophisticated ab initio calculations.\cite{voth3}

At each step of the Molecular Dynamics runs, the simulation procedure 
involved the construction of a connectivity pattern of hydrogen bonds (HB), 
linking the water molecule exhibiting strongest H$_3$O$^+$ character 
(hereafter referred to as the pivot water and denoted H$_2$O$^*$) with its 
first, second -- and eventually -- third solvation shells.  This pattern allowed 
the identification of the instantaneous diabatic states included in the 
construction of $\hat{\rm H}_{{\rm EVB}}$.  For $n=10$ and $n=20$, the number 
of EVB states considered was intermediate between 8 and 15;  this number 
went up to typically 20, for simulations of the largest aggregates.

The dynamics of the classical nuclei was generated from the following Newton's 
equations of motions:
\begin{equation}
\label{eq2}
M_k \frac{{\rm d}^2 {\bf R}_k }{{\rm d}t^2 } = - \nabla_{{\bf R}_k}
\epsilon_0\left(\{{\bf R}\}\right) \ \  ;
\end{equation}
where $\epsilon_0\left(\{{\bf R}\}\right)$ represents the 
lowest eigenvalue of the EVB Hamiltonian, namely:
\begin{equation}
\hat{\rm H}_{{\rm EVB}} |\psi_0\rangle = \epsilon_0(\{{\bf R}\})
| \psi_0\rangle  \ \ . 
\end{equation}
In the previous equation, $| \psi_0\rangle $ is the ground state eigenstate of 
$\hat{\rm H}_{{\rm EVB}}$, whose expression in terms of the diabatic states can
be written as:
\begin{equation}
|\psi_0\rangle =  \sum_i c_i |\phi_i\rangle \ \ .
\end{equation}
The index of the largest expansion coefficient, $c_i$, identifies
the instantaneous pivot water, which can be eventually  updated, in the advent 
of a proton translocation episode.  

Diagonal elements $h_{ii} (\{{\bf R}\})$ included inter- and  intra-molecular
interactions involving the  H$_3$O$^+$ group and the rest of the
water molecules,\cite{voth3} that were modeled using the flexible TIP3 
model.\cite{tip3p}
All simulation experiments began with a $\sim 0.5$ ns preliminary thermalization stage,
at  temperatures close to $T\sim 200$ K. In this thermal regime, the 
clusters exhibited structural and dynamical characteristics similar to those observed
in liquidlike phases. Subsequent to this equilibration period, statistics were 
collected from constant energy trajectories, lasting typically $\sim 5-10$ ns. 

\section{Results}
\subsection{Solvation structures}

Mesoscopic clusters represent inhomogeneous environments at the nanometer scale. 
Perhaps the first important question to be answered in connection with the solvation
of an excess proton in a water cluster concerns the average localization of the 
excess charge. In this context, the consideration of 
two local density fields with respect to the center of mass of the 
cluster will be useful: The first one, $\rho_{\rm CM-O^*}(r)$, involves the 
position of the pivot water and is defined as:
\begin{equation}
\rho_{\rm CM-O^*}(r) = \frac{1}{4 \pi r^2 }  \langle \delta (|
{\bf r}_{\rm O^*}- {\bf R}_{\rm CM} | - r )\rangle  \ . 
\end{equation}
Note that $\rho_{\rm CM-O^*}(r)$ represents the probability distribution of 
finding the pivot at a distance $r$ from the center of mass of the aggregate. 
In the previous equation, $ {\bf r}_{\rm O^*}$ and ${\bf R}_{\rm CM}$  denote the 
positions of the pivot and the center of mass of the cluster, respectively. 
The second density, $\rho_{\rm CM-W}(r)$, involves the rest of the water molecules and 
is defined as:
\begin{equation}
\rho_{\rm CM-W}(r) = \frac{1}{4 \pi r^2  } \sum_i^{n-1}   \langle \delta (|
{\bf r}_{i=1}^{\rm o}- {\bf R}_{\rm CM} | - r )\rangle  \ \ ; 
\end{equation}
where ${\bf r}_i^{\rm o}$ represents the coordinate of oxygen site in the 
$i$-th molecule and the sum involves all the water molecules, except the one 
acting as pivot.

Results for both functions are depicted in Fig. 1.  The profiles of the 
bottom panel contrast sharply to those shown in the upper ones: For the smallest 
size considered, $n=10$, it is evident that the excess charge localizes 
closer to the center of mass of the aggregate.  A snapshot of a typical cluster 
configuration for a protonated clusters of this size is depicted Fig. 2.  
Note that the overall structure looks as a mostly planar, star-like, 
arrangement of hydrogen bonded water molecules, with the three-coordinated 
hydronium lying at a central position.

The situation changes at a qualitative level  in larger 
aggregates (see middle and top panels of Figs. 1).
Under these circumstances, the water local density at the central part 
looks much more uniform and its value does not differ substantially from the 
usual one for  bulk water at ambient conditions: 
$\rho_{\rm CM-O^*}(r\sim 0)\sim \rho^{bulk}_w=0.033$ \AA$^{-3}$. 
Snapshots for cluster  configurations in this size range are also included 
in Fig. 2 and reveal that, in both cases, the aggregates 
exhibit a three-dimensional hydrogen-bonded structure, in which it is possible 
to discriminate bulk from surface states environments. As for the position of the proton 
is concerned, there are clear indications that the regions for stable proton 
solvation in both cases are located at the corresponding cluster boundaries. 
This observation has been previously reported\cite{cl}  and has also been found in
simulations of protons at macroscopic water-air interfaces.\cite{interfaces} 

It is also of interest to investigate the closest solvation structures 
of the excess charge. In bulk water, the microscopic description of such 
environments includes a whole series of arrangements which are intermediate 
between two limiting moieties, with well defined geometry: We are referring to 
the symmetric Zundel (ZDL) dimer\cite{zundel} [H$_2$O-H-OH$_2$]$^+$ and 
three-coordinated Eigen (EGN) cation\cite{eigen} [H$_3$O$\cdot$(H$_2$O)$_3$]$^+$. 
Interconversions between these structures take place in the picosecond time 
scale and seems to be triggered by changes in the intramolecular connectivity 
between the proton closest solvation shells. The relative prevalence of these 
structures can be conveniently monitored by considering probability 
densities associated to an appropriate order parameter identifying EGN and 
ZDL complexes.  In the MS-EVB model, such order parameter can be expressed in 
terms of the two largest expansion coefficients (see Eq. 3) as:\cite{voth3,voth5}
\begin {equation}
\ \ \xi=c_1^2-c_2^2 \ \ \ .
\end {equation}
ZDL and EGN solvation environments are characterized by values of $\xi$ close 
to 0 and 1, respectively. 
In Fig. 3 we present results for $W({\xi})$, the free energy associated to 
the order parameter:
\begin{equation}
\ \ \beta W(\tilde{\xi})  \propto - \ln
\langle \delta(\xi - \tilde{\xi})\rangle \ \ ;
\end{equation}
where $\beta$ is the inverse of Boltzmann constant times the
temperature and  $\langle .... \rangle$ represents a statistical average.
For the sake of comparison, we have also included in the figure the results for 
bulk water reported in Refs. [14, 16]  In all cases, the most stable structures -- which
correspond to the global minimum of $W(\xi)$ -- are characterized
by values of $\xi_{min}\sim 0.55$. The magnitude of the
free energy differences between EGN-like and ZDL-like structures though,
does depend on the particular environment considered. Results for $W(\xi=0)$
in aggregates with $n= 21$ and $n= 125$ are practically
identical, revealing that the characteristics of the surface dynamical modes 
triggering EGN-ZDL interconversions are comparable for both cluster sizes. 
On the other hand, for $n= 10$, EGN-ZDL interconversions become much more 
frequent and are dominated exclusively by polarization fluctuations in 
the aggregate. The differences in temperature preclude the direct comparison 
of these results to those observed in bulk water at ambient conditions; 
anyhow, our simulations show that the free energy barrier for the isotropic 
bulk phases is $\sim 0.5$ kcal mol$^{-1}$ higher than the one observed at the 
surface of the  largest clusters.

To gain additional insight on the local coordination of the excess charge,
we also examined pair correlations involving the O$^*$ site. In the top panel of 
Fig. 4 we show plots for pair correlations between the pivot O$^*$ and 
oxygen (top panel) and hydrogen (bottom panel) sites in the rest of the 
aqueous domains.  Correlations with oxygen sites are dominated by main 
peaks located at $r\sim 2.5$ \AA \ which involves $\sim 2.9$ water-oxygen 
sites acting as HB acceptors. The magnitude and position of these peaks 
do not change with the cluster size and are similar to those observed in 
bulk water.\cite{voth3} 
The overall shapes of the pivot-hydrogen profiles are also dominated by 
main peaks shifted $\sim 0.6$ \AA \ outward and include six hydrogen atoms 
belonging to water molecules in the pivot first solvation shell. 
Consequently, we are led to conclude that, in all aqueous environments 
investigated, the closest solvation structures of the proton are preserved,
being characterized by a first solvation shell composed by three water 
molecules, acting as HB acceptors with  no evidence of pivot 
HB of the type O-H$\cdots$O$^*$ .

\subsection{Proton transfer}

Having established the main features of the equilibrium solvation 
structures of the proton, we now move to the consideration of the 
dynamical characteristics related to transfer processes in aqueous 
clusters.  In order to acquire a preliminary notion of the timescales 
characterizing proton translocation events in water clusters, in Fig. 5 
we present the time evolutions of the pivot label during a 1 ns time 
interval. 
At a first glance, it is selfevident that proton jumps become much more 
frequent as we move to larger aggregates. At $T\sim 200$ K 
and for $n=10$, the dynamics of the pivot label can be pictured as a 
sequence of three well differentiated classes of episodes: $(i)$ on one hand, 
one observes fast resonances in the subpicosecond time domain, between ZDL
pairs interrupted by $(ii)$ stages during which the pivot label remains
practically unchanged for several hundreds of picoseconds. 
$(iii)$ The stabilization of the proton in a different water pivot is a 
much more rare event since it is normally preceded by major rearrangements 
of the overall cluster geometry. These modifications involve changes 
in the localization of the potential donor and acceptor molecules, requiring
the migration of the acceptor molecule towards a more central position 
within the aggregate. This is operated by the breakage and regeneration 
of new HB, that take place over lengthscales comparable to the overall size 
of the aggregate.  
Although the time evolutions of the pivot label for larger aggregates present 
similar qualitative characteristics, translocation episodes do occur 
much more frequently.  
Concerning possible mechanisms that might trigger transfer episodes at
cluster surfaces, 
the direct inspection of the close vicinity of the acceptor-donor pair 
revealed that the transfer dynamics
is operated by subtle changes in the HB structure in the second and third
solvation shells. In this respect, although we have not performed a detailed
trajectory analysis of reactive processes, we are led to believe that the 
surface mechanisms that regulate translocations resemble very much those 
observed in bulk macroscopic phases.\cite{mechanism}

A more precise estimate for the proton transfer rates can be 
extracted from population relaxation time correlation functions of the type:
\begin{equation}
\label{ct}
C(t) = \frac{ \langle
\delta h_i(t)  . \delta h_i(0) \rangle
}{ \langle
(\delta h_i)^2 \rangle } \ \ \ ;
\end{equation}
where $\delta h_i(t) = h_i(t) - \langle h_i \rangle$ denotes the 
instantaneous fluctuation of
the population of the $i$th reactant away from its equilibrium value. In Eq. 
\ref{ct},  the
dynamical variable $h_i(t)$ is
unity if the proton is localized in the $i$-th diabatic state at time $t$ 
and zero otherwise. 
Provided Onsager's regression hypothesis\cite{dc} remains valid, the 
exponential decay at long times of $C(t)$ should yield an estimate of the proton 
transfer rate, $\tau_{tr}^{-1}$. Plots for $\ln C(t)$ are presented in Fig. 5; after 
fast transients ascribed to ZDL resonances, the three curves present
single exponential decays.  Compared to the bulk result
$\tau_{tr}^{blk}\sim 2-4$ ps, the estimates for the characteristic timescales 
in  clusters are intermediate between one and two orders of magnitude larger:
$\tau_{tr}^{125}\sim 20$ ps and $\tau_{tr}^{10}\sim 130$ ps.

\section{Concluding remarks}

The structural and dynamical characteristics of excess protons embedded in 
liquid-state nanoclusters of the type H$^+$[H$_2$O]$n$ described in this 
paper illustrate  distinctive behaviors, depending on the cluster size considered. More
specifically, we focussed attention to $n=10$, 21  and 125.
Taking into consideration that the cluster radii scale as $R \propto n^{1/3}$, 
we tried to cover different scenarios: on one hand, we considered  $n \alt 20$  aggregates, 
in which it is impossible to discern surfance from bulk states;  on the
other, $n\sim 100$ clusters -- with $R$ of the order of three molecular diameters -- 
perhaps the smallest ones in which such distinction can be clearly established.

In small
aggregates, the overall structures of H$^+$[H$_2$O]$_{10}$ clusters can be pictured 
as dendritic-like arrays of hydrogen bonded water molecules, with the excess proton 
occupying a central position. At larger sizes, there is a clear
tendency for surface solvation of the excess charge. 
Note at in all cases, these two solvation structures preserve
two important characteristics of the closest aqueous environments  
of the excess charge: $(i)$ the three-coordinated, HB donor structure of 
hydronium with respect to its first solvation shell and $(ii)$ the absence of HB acceptor
structure of the type O-H$\cdots$O$^*$. We remark that this anisotropies 
have also been reported in previous cluster studies,\cite{cl,cl1} at water-air 
interfaces,\cite{interfaces} and confined water within membranes.\cite{nafion}

The observed proton dynamics dynamics in small clusters can be pictured as a sequence of 
three different episodes: $(i)$ First, one observes resonance stages, during which the 
proton seems to be delocalized over a tagged ZDL pair. These fast dynamical
modes of the pivot label seem to be dictated by polarization fluctuations originated 
in the closes solvation environments of the excess charge; 
$(ii)$ The latter fluctuations seem to be also the controlling agent for the localization
stages, during which the resonances cease and the proton gets localized in a single
water molecule, for periods that may last typically $\sim 100$ ps;
$(iii)$ Finally, one also observes much rare proton translocation episodes, which are
preceded by global rearrangements of the hydrogen bond connectivity of the
the cluster. These dynamical characteristics are maintained at a qualitative 
level for  $n=21$ and $n=125$ aggregates, although the translocations become much
more frequent and the triggering mechanisms require much more subtle modifications
in the HB connectivity that those observed in smaller aggregates.

\vspace{1cm}

\section{Acknowledgements}
DL and JR are staff members of CONICET-Argentina.

\newpage

\begin{figure}[thb]
\epsfysize=4.in
\epsfbox{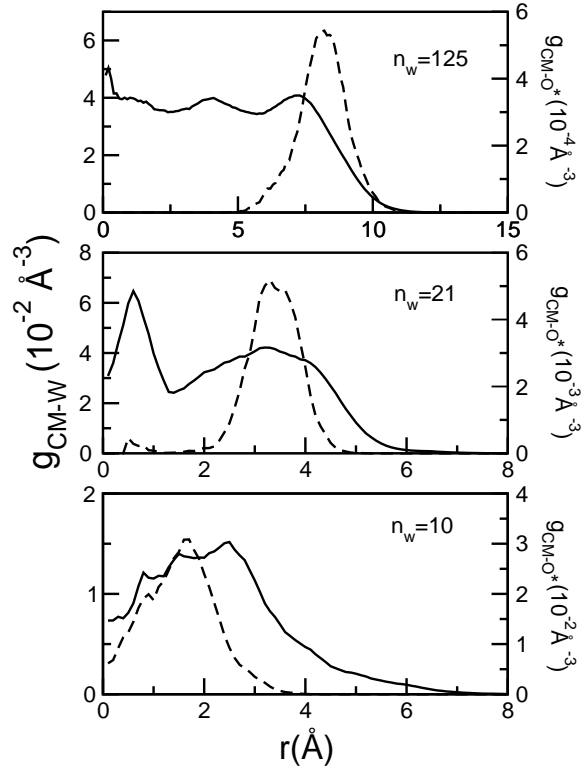}
\caption{Local density fields with respect to the center of mass of 
aqueous clusters of different sizes: water oxygen (solid line, left axis); 
pivot oxygen (dashed line, right axis).
}
\end{figure}
\newpage

\begin{figure}[thb]
\epsfysize=4.in
\epsfbox{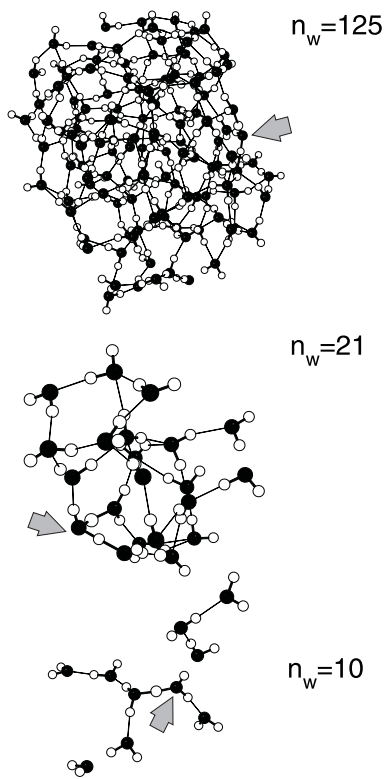}
\caption{Snapshots for typical configurations of protonated water clusters. The
arrows indicate the instantaneous position of the pivot oxygen}
\end{figure}
\newpage

\begin{figure}[thb]
\epsfysize=3.5in
\epsfbox{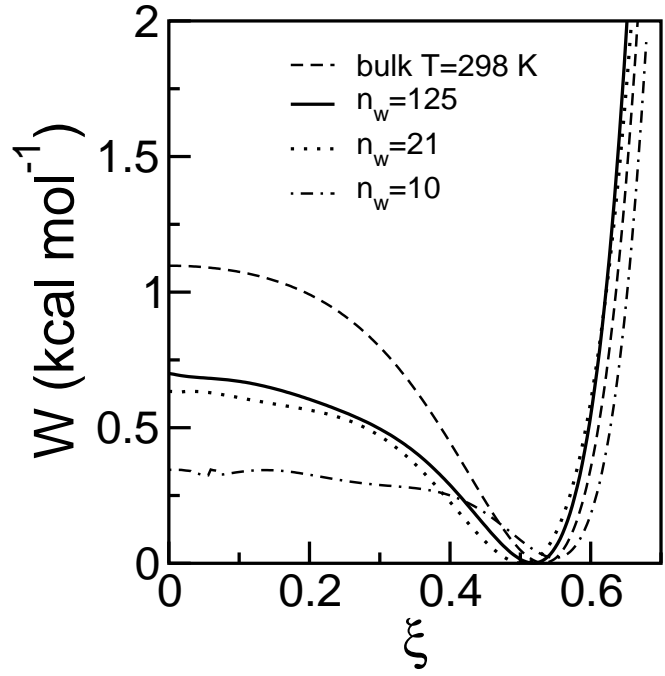}
\caption{
Free energy associated to the asymmetric order parameter
$\xi$ describing Eigen-Zundel interconversions  in protonated aqueous clusters and 
bulk water at ambient conditions.
The curves were brought to the same reference value at their corresponding minima.
}
\end{figure}
.

\newpage

\begin{figure}[thb]
\epsfysize=3.in
\epsfbox{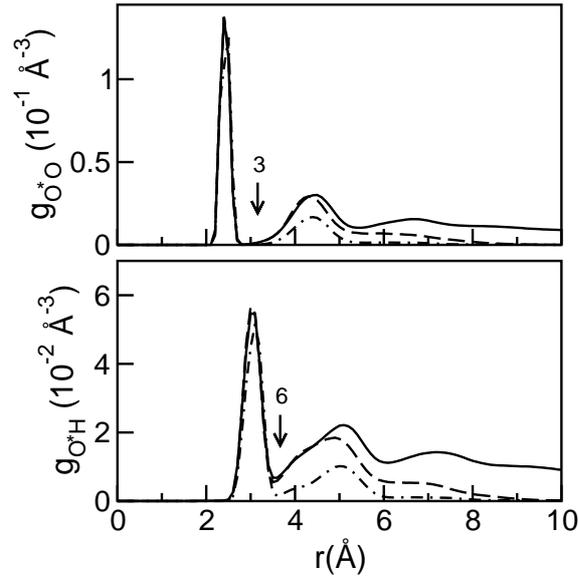}
\caption{
Pair correlation functions between 
the pivot and oxygen (top panel) and hydrogen (bottom panel) sites in water clusters of
different sizes. Solid line: $n=125$ (solid line); dashed line: $n=21$;
dot-dashed line: $n=10$. The arrows indicate the are under the main peaks.
}
\end{figure}

\newpage

\begin{figure}[thb]
\epsfysize=3.in
\epsfbox{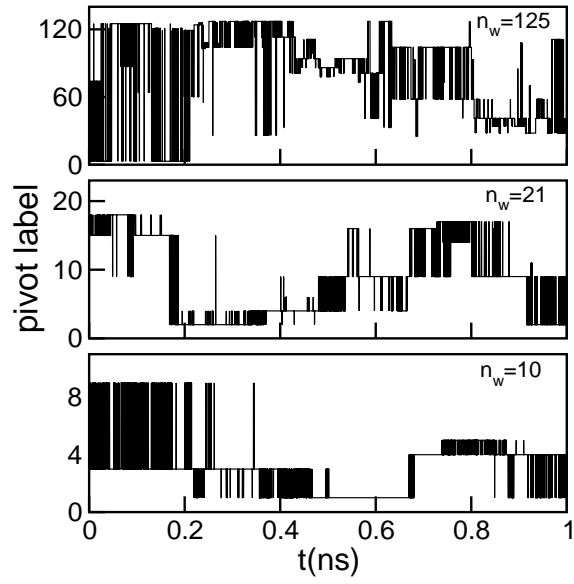}
\caption{Time evolutions of the pivot label in aqueous cluster of different sizes.}
\end{figure}

\newpage

\begin{figure}[thb]
\epsfysize=3in
\epsfbox{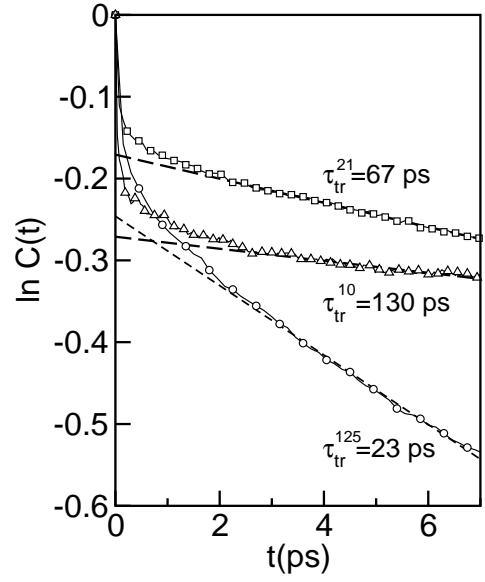}
\caption{Logarigthm of the population relaxations for the pivot 
label in aqueous clusters of different sizes. 
$n=10$: triangles; $n=21$: squares;  $n=125$: circles.
The characteristic times for the transfer $\tau_{tr}$ were obtained 
from the inverse of the corresponding slopes at long times (dashed lines).}
\end{figure}

\end{document}